\documentclass[aps,prc,twocolumn,superscriptaddress,showpacs,showkeys]{revtex4-1}
\usepackage{graphicx}
\begin{document}
\title{Search for axioelectric effect of solar axions using BGO scintillating bolometer}

\author{A.V.~Derbin$^1$, L.~Gironi$^{2,3}$, S.S.~Nagorny$^{4,5}$, L.~Pattavina$^4$, J.W.~Beeman$^6$,
F.~Bellini$^{7,8}$, M.~Biassoni$^{2,3}$, S.~Capelli$^{2,3}$, M.~Clemenza$^{2,3}$, I.S.~Drachnev$^{1,5}$,
E.~Ferri$^{2,3}$, A.~Giachero$^{2,3}$, C.~Gotti$^{2,3}$, {A.S. Kayunov$^1$}, C.~Maiano$^{2,3}$, M.~Maino$^{2,3}$,
V.N.~Muratova$^1$, M.~Pavan$^{2,3}$, S.~Pirro$^4$, D.A.~Semenov$^1$, M.~Sisti$^{2,3}$, E.V.~Unzhakov$^1$.}

\affiliation{St. Petersburg Nuclear Physics Institute, Gatchina 188350 - Russia\\
$^2$INFN - Sezione di Milano Bicocca, Milano I-20126 - Italy\\
$^3$Dipartimento di Fisica, Universit\`a di Milano-Bicocca, Milano I-20126 - Italy\\
$^4$INFN - Laboratori Nazionali del Gran Sasso, Assergi (L'Aquila) I-67100 - Italy\\
$^5$Gran Sasso Science Institute, INFN, L'Aquila (AQ) I-67100 - Italy\\
$^6$Lawrence Berkeley National Laboratory, Berkeley, California 94720 - USA\\
$^7$INFN - Sezione di Roma, Roma I-00185 - Italy\\
$^8$Dipartimento di Fisica - Universit\`a di Roma La Sapienza, Roma I-00185 - Italy}

\begin{abstract}
A search for axioelectric absorption of solar axions produced in the $p + d \rightarrow {^3\rm{He}}+\gamma~(5.5~
\rm{MeV})$ reaction has been performed with a BGO detector placed in a low-background setup. A model-independent limit
on the combination of axion-nucleon and axion-electron coupling constants has been obtained: $| g_{Ae}\times g_{AN}^3|<
1.9\times 10^{-10}$ for 90\% confidence level. The constraint of the axion-electron coupling constant has been obtained
for hadronic axion with masses of (0.1 - 1) MeV: $|g_{Ae}| \leq (0.96 - 8.2)\times 10^{-8}$.
\end{abstract}



\maketitle

\section{INTRODUCTION}

In 1977 Peccei and Quinn suggested a new approach for the solution of the strong CP problem \cite{Pec77}. They
introduced a new global chiral symmetry U(1) which was spontaneously violated at some energy $f_A$ and compensated the
CP-invariant term of the QCD Lagrangian. Later in 1978 Weinberg \cite{Wei78} and Wilczek \cite{Wil78} showed that such
spontaneous breaking should lead to the appearance of the new neutral pseudoscalar particle - an axion. The initial
hypothesis (Weinberg-Wilczek-Peccei-Quinn axion) expected the U(1) symmetry to be broken at the electro-weak scale $f_A
= 250$ GeV and yielded certain predictions for the values of axion coupling constants with photons ($ g_{A\gamma}$),
electrons ($ g_{Ae}$) and nucleons ($ g_{AN}$) and also the axion mass ($m_A$). The following series of reactor and
accelerator experiments disproved this initial hypothesis \cite{PDG12}.

New axion models removed the restrictions on the $f_A$ value, allowing it to be extended up to the Planck mass
$m_P\approx 10^{19}$ GeV. These models can be separated in two major classes: hadronic (or
Kim-Shifman-Vainstein-Zakharov models) \cite{Kim79,Shi80} and GUT (or Dine-Fischler-Srednicki-Zhitnitskii models)
\cite{Zhi80,Din81}. In both cases the axion coupling to ordinary particles is suppressed by the energy scale that
characterizes the symmetry breaking, i.e., the axion decay constant $f_A$. The axion mass is given in terms of neutral
pion properties:
\begin{equation}\label{ma}
  m_A\approx (f_\pi m_\pi /f_A) (\sqrt{z}/(1+z)),
\end{equation}
where $m_\pi$ and $f_\pi$ are, respectively, the mass and decay constant of the $\pi^0$ meson and  $z = m_u/m_d$ is $u$
and $d$ quark-mass ratio. Taking those values into account  one can present equation (\ref{ma}) as:
$m_A(\rm{eV})\approx 6.0 \times 10^6/\it{f_A} {\rm{(GeV)}}$. In new models, due to the lack of upper limit on $f_A$,
the axion interactions with ordinary matter appear to be significantly suppressed, so these axions were labeled as
"invisible".

Various laboratory experiments as well as astrophysical and cosmological arguments have been used to constrain the
allowed range for $f_A$ or, equivalently, for the axion mass $m_A$.  The most stringent upper limit on the axion mass
derived from the astrophysics is $m_A < 0.01 \rm{~eV}$ \cite{Jan96,Lei14}, while cosmological arguments yield the lower
limit of $m_A > 10^{-5} \rm{~eV}$ \cite{Abb83}.

Thus, the majority of the experimental axion searches examine the mass range of $10^{-6}$ to $10^{-2}$ eV and relic
axions of these masses are considered to be very favorable candidates for cold dark matter particles. It should be
noted though, that experimental upper limits on the axion mass are obtained from the restrictions on the axion coupling
constants $g_{A\gamma}$, $g_{Ae}$, and $g_{AN}$ and are strongly model dependent.

Direct laboratory searches for solar axions with CAST and IAXO helioscopes \cite{Ari14,Arm14} and relic axions with
ADMX haloscope \cite{Sho14,Bib13} rely on the axion-two-photon vertex, allowing for axion-photon conversion in external
electric or magnetic fields \cite{PDG12}. Reactions of axioelectric effect in atoms and resonant absorption by nuclei
can be induced by the axion-electron and axion-nucleon couplings \cite{Der13A}.

One may consider more general axion-like particles (ALPs), where the axion coupling constants and the axion mass are
independent. Possible examples of ALPs that have already been studied include light CP-odd Higgs bosons
\cite{Hoo09,And10A}. Similarly, light spin 1 particles called hidden sector photons or light minicharged particles
occur in case of embedding the standard model into the string theory. Several experiments have explored this more
general region. The current constraints are compiled in \cite{Bak13,Jae12}.

In addition, there are suggestions for strong CP problem solution, which allow the existence of axions with quite a
large mass (~1 MeV), while their interaction with ordinary particles remain at the level of the invisible axions. The
models rely on the hypothesis of a world of mirror particles \cite{Ber01} and SUSY at the TeV scale \cite{Hal04}. The
existence of these heavy axions is not precluded by the laboratory experiments or astrophysical data.

 This article describes the experimental search for 5.5 MeV axions, performed with the use of $\rm{Bi_4Ge_3O_{12}}$
(BGO) bolometric detectors. Solar axions of the corresponding energy can be produced by $p + d \rightarrow\rm{^3He}+ A$
reaction. Their flux should be proportional to the $pp$-neutrino flux, which has been estimated with high degree of
accuracy in \cite{Ser09}. The incident solar axions are supposed to interact with BGO crystal via the reaction of
axioelectric effect ${\rm A}+e+Z\rightarrow e+Z$. This kind of interaction is governed by axion-electron coupling
constant $g_{Ae}$ and its cross section depends on the charge of nucleus as $Z^5$. From this point of view the BGO
detector is a very suitable target, because of the high Z value of bismuth nucleus ($Z_{Bi} = 83$).

Recently, the existence of the high energy solar axions and axions from a nuclear reactor have been investigated by the
Borexino \cite{Bel08,Bel12}, the CAST \cite{And10} and the Texono \cite{Cha07} collaborations. The search for 5.5 MeV
axions with BGO scintillating detectors have been performed in \cite{Der10,Der13}.

\section{ AXION PRODUCTION IN NUCLEAR MAGNETIC TRANSITIONS AND THE AXIOELECTRIC EFFECT}
The axions of the ~MeV energy region can be produced by reactions of main solar cycle and CNO chain. The $p + d
\rightarrow {^3\rm{He}} + \gamma$ reaction is expected to make the dominant contribution to the total axion flux. In
this case 5.5 MeV axion is emitted instead of $\gamma$-quantum.

The standard solar model implies that 99.7\% of the total amount of solar deuterium are produced by the two-proton
fusion: $p + p \rightarrow d + e^+ + \nu_e$ and the rest 0.3\% are formed as a result of  $p+ p + e^- \rightarrow  d +
\nu_e$ reaction. The estimate for solar axion flux can be obtained from the value of the $pp$-neutrino flux: $6.0\times
10^{10} {\rm{cm}}^{-2} {\rm{s}}^{-1}$ \cite{Ser09}. The flux ratio between axions and neutrinos depends on the
axion-nucleon coupling constant $g_{AN}$, which includes the isoscalar $g^0_{AN}$ and isovector $g^3_{AN}$ components.

In order to produce an axion, the nuclear transition has to be of magnetic type. In case of the
$p(d,{^3\rm{He}})\gamma$ reaction it corresponds to the proton capture with a zero orbital momentum. The probability of
the proton capture from the $S$ state for energies below 80 keV has been measured in \cite{Sch97}. For ~1 keV protons
the M1-fraction of the total $p(d,{^3\rm{He}})\gamma$ cross section is equal to $\chi$ = 0.55. Since the proton capture
from the $S$ state corresponds to the isovector transition, the axion production probability ratio $(\omega_A /
\omega_\gamma)$ will depend only on the axion-nucleon coupling constant $g^3_{AN}$
\cite{Don78,Raf82,Avi88,Hax91,Der97}:
\begin{equation}\label{ratio}
\frac{\omega_{A}}{\omega_{\gamma}} =
 \frac{\chi}{2\pi\alpha}\left[\frac{g_{AN}^{3}}{\mu_3}\right]^2\left(\frac{p_A}{p_\gamma}\right)^3 = 0.54(g_{AN}^{3})^2
 \left(\frac{p_A}{p_\gamma}\right)^3.
\end{equation}
where $p_{\gamma}$ and $p_{A}$ are, respectively, the photon and axion momenta; $\alpha\approx 1/137$ is the
fine-structure constant;  and $\mu_3 = \mu_p - \mu_n \approx 4.71$ is isovector nuclear magnetic momenta.

In case of hadronic axion model, the coupling constant $g^3_{AN}$ can be expressed in terms of the axion mass
\cite{Sre85, Kap85}:
\begin{equation}\label{gan3}
g_{AN}^{3}=-2.75 \times 10^{-8}(m_A/1 {\rm{eV}}).
\end{equation}

In the GUT-axion model $g^3_{AN}$ constant contains additional unknown parameter $cos^2\beta$, though its value is
still of the same order of magnitude. The ratio is ($0.3 - 1.5$) in comparison with $g^3_{AN}$ value for hadronic model
\cite{Sre85}.
\begin{figure}
\includegraphics[width=9cm,height=10.5cm]{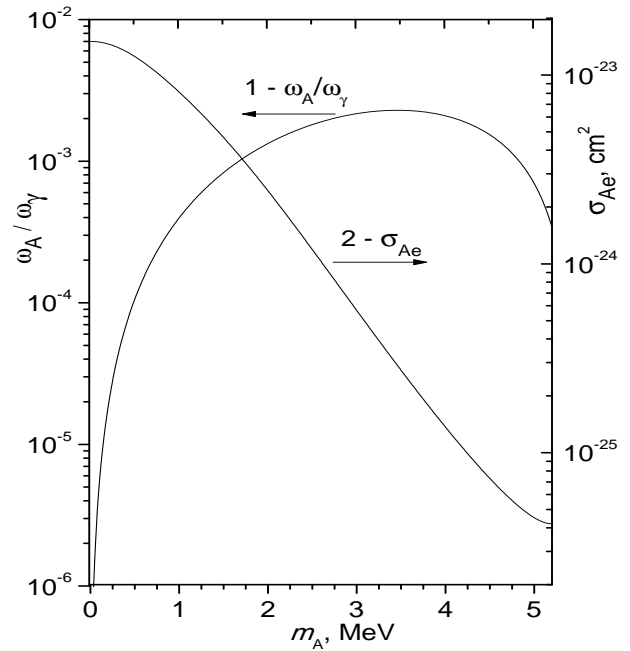}
\caption {The ratio ($\omega_A/\omega_\gamma$) in the $p + d \rightarrow {^3{\rm{He}}} + \gamma$ reaction (curve 1,
left-hand scale); the cross section $\sigma_{Ae}$ for 5.5-MeV axions on Bi-atoms for $g_{Ae} = 1$ (curve 2, right-hand
scale). } \label{Fig1}
\end{figure}
Fig. \ref{Fig1} shows the calculated $\omega_A/\omega_\gamma$ ratio value versus the axion mass $m_A$. At the Earth's
surface the axion flux amounts to:
\begin{eqnarray}\label{FluxA}
\Phi_A = \Phi_{\nu p p}(\omega_A/\omega_\gamma)
\end{eqnarray}
where $\Phi_{\nu p p} = 6.0 \times 10^{10} {\rm{cm}}^{-2} {\rm{s}}^{-1}$ is the $pp$-neutrino flux.

In order to detect 5.5 MeV solar axions, we considered the reaction of axioelectric effect $A + Z + e \rightarrow Z +
e$,  which is governed by the axion-electron coupling constant $g_{Ae}$. The cross section of this process depends on
the charge of the nucleus as $Z^5$, thus making materials with high $Z$ values favorable for use in such experiments.
For bismuth atoms, the cross section of axioelectric effect exceeds the one for Compton conversion by almost two orders
of magnitude. Taking into account the fact that electron detection efficiency is close to 100\% and the background at
5.5 MeV does not overlap with the regions of natural radioactivity, the resulting sensitivity to $g_{Ae}$ and $g_{AN}$
remains sufficient even if the target mass is relatively low.

When the axioelectric effect occurs, the atom emits an electron with energy of $E_e=E_A - E_b$, where $E_b$ is the
electron binding energy (similar to photoelectric effect). The cross section for the axioelectric effect was calculated
in \cite{Zhi79}, assuming that that $E_A\gg E_b$ and $Z\ll 137$.
\begin{center}
\begin{eqnarray}
\nonumber \sigma_{Ae} = 2(Z\alpha m_e)^5\frac{g^2_{Ae}}{m_e^2}\frac{p_e} {p_A}\ [\frac{4E_A(E^2_A+m^2_A)}{(p^2_A- p^2_e)^4}-\frac{2E_A}{(p^2_A -p^2_e)^3} \\
\nonumber-\frac{64}{3}p^2_ep^2_Am_e\frac{m^2_A}{(p^2_A -p^2_e)^6}-\frac{16m^2_Ap^2_AE_e}{(p^2_A-p^2_e)^5}-~~\\
-\frac{E_A}{p_ep_A}\frac{1}{(p^2_A-p^2_e)^2}\ln\frac{p_e+p_A}{p_e-p_A}].~~~\label{sigmaAE}
\end{eqnarray}
\end{center}

Fig. 1 contains the dependence of cross section on the axion mass, in assumption that $g_{Ae} = 1$. The most
significant contribution to the total cross section is made by K-shell electrons. The rest of the electrons was taken
into account by introducing a factor $5/4$, same as in case of the photoelectric effect.

There are two requirements for the flux of 5.5 MeV axions to be proportional to the $pp$-neutrino flux at the surface
of Earth. The axion lifetime has to exceed the time of flight between Sun and Earth and also axion flux should not be
considerably reduced due to absorbtion by solar matter. From these conditions one can obtain limits for values of axion
coupling constants that are available for terestrial experiments \cite{Raf82,Bel08,Der10,Der13,Bel12}.

In order to reach the solar surface from the center of the Sun axions have to pierce through $\approx 6.8\times
10^{35}$ electrons and $\approx 5\times 10^{35}$ protons per ${\rm{cm}}^{-2}$ of solar matter. The sensitivity of
terrestrial experiments to $g_{Ae}$ constant is strictly limited by Compton conversion of axion into photon. The cross
section of this reaction for 5.5 MeV axions depends weakly on the axion mass and can be written as: $\sigma_{cc}\approx
g^2_{Ae}4\times 10^{-25} {\rm{cm}}^2$ \cite{Don78,Bel12}. For $|g_{Ae}|$ values below $10^{-6}$ axion flux remains
almost unaffected by absorption.

The $g_{A\gamma}$ limits imposed by axion-photon interaction are obtained from the reaction of axion conversion  inside
magnetic field of an atomic nucleus. Taking the known proton and $^4\rm{He}$ densities into account, one can see that
axions will be able to effectively escape the Sun if $|g_{A\gamma}| < 10^{-4}~{\rm{GeV}}^{-1}$.

The coupling of axions with atomic nuclei ($g_{AN}$) leads to reduction of axion flux due to the reaction of
photo-dissociation. As it was shown in \cite{Raf82} the dominant contribution for this process is made by the following
reaction: ${A + {^{17}\rm{O}}\rightarrow {^{16}\rm{O}}+n}$. Provided the value of nuclear coupling constant is
$|g_{AN}^3-g_{AN}^0|< 10^{-2}$, the resulting axion flux change does not exceed 10\%. Since the axion absorption
produced by isovector transition $A(^3\rm{He},d)p$ is negligible, the range of available $|g_{AN}^3|$ values remains
arbitrary.

For the masses on an axion exceeding $2m_e$, the primary decay mode should be the production of an electron-positron
pair. Assuming that 90\% of the total amount of produced solar axions reach the Earth, we can set the sensitivity upper
limit for the electron coupling constant as $|g_{Ae}| < (10^{-12}-10^{-11})$ \cite{Der10}. In case of the axion masses
below $2m_e$, $e^+ + e^-$ option is no longer available, but decay into two $\gamma$-quanta is still valid. The value
of decay probability is determined by the axion-photon coupling constant $g_{A\gamma}$ and the axion mass $m_A$:
$\tau_{A\rightarrow\gamma\gamma}=64\pi/g^2_{A\gamma}m^3_A$. Current experimental constraints ($g_{A\gamma} < 10^{-9}$
${\rm{GeV}}^{-1}$) yield $\tau_{cm} = 10^5$ in case of 1 MeV axions. It suggests that axion flux remains practically
unaffected by $A\rightarrow 2\gamma$ decays, even for 5 MeV axions.

\section{EXPERIMENTAL SETUP}
The detector used for this study is an array of ${\rm{Bi}}_4{\rm{Ge}}_3{\rm{O}}_{12}$ (BGO) scintillating bolometers
\cite{Car12}, containing 1.65 kg of Bi. Four cubic ($5\times5\times5$ $\rm{cm}^3$) BGO crystals, with all optical faces
were arranged in a four-plex module, one single plane set-up. The scintillation light produce by particle interaction
in the BGO absorbers was monitored with an auxiliary bolometer made of high-purity germanium, operated as a light
detector (LD) \cite{Bee13}.

The detector was installed in the $^3\rm{He}/^4\rm{He}$ dilution refrigerator in the Hall C of the underground
laboratory of L.N.G.S. ($\simeq$ 3650 m w.e.) and operated at a temperature of few mK. The four crystals and the LD
were housed in a highly pure copper structure, the same described in \cite{Ale12}. Given the large light yield (LY) of
BGO crystals \cite{Ort11} a single wide area LD was faced to the entire array and no reflecting foil was used. This
type of assembly did not prevent us from exploiting the powerful particle discrimination capability of this
scintillating bolometer. The four BGO crystals were analyzed to search for the 5.5 MeV axions.

Coupled to each bolometer there is a Neutron Transmutation Doped (NTD) germanium thermistor that acts as a thermometer:
recording the temperature rises produced by particle interaction in the absorbers and producing voltage pulses
proportional to the energy deposition. These pulses then are amplified and fed into an 18-bit analog-to-digital
converter. Software triggers ensure that every thermistor pulse is recorded. Details on our electronics and on the
cryogenic set-up can be found elsewhere \cite{Pir00,Arn06}.

The amplitude and the shape of the pulses is then determined by the off-line analysis. To maximize the signal-to-noise
ratio, the pulse amplitude is estimated by means of the Optimum Filter (OF) technique \cite{Gat86,Rad67}.

The heat and light channels were energy-calibrated by means of gamma ($^{40}\rm{K}$ and $^{232}\rm{Th}$) and X-ray
($^{55}\rm{Fe}$) sources, respectively. The relation between pulse amplitude and energy was parameterized with a first
order polynomial fit.
\begin{figure}
\includegraphics[width=9cm,height=10.5cm]{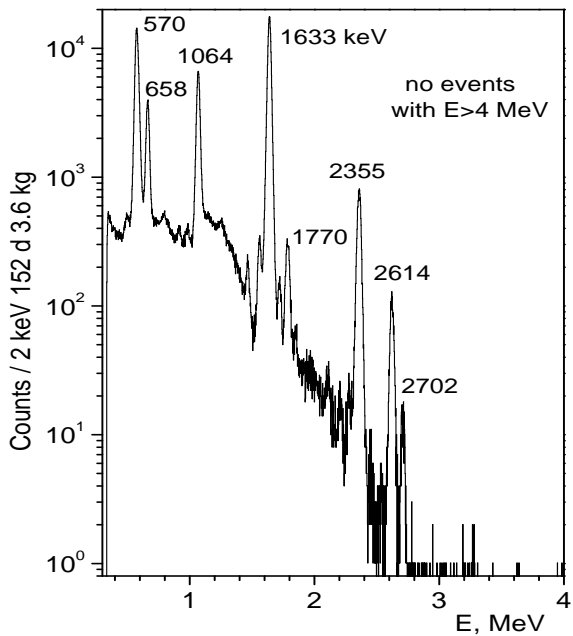}
\caption {The energy spectrum of the four BGO detectors ($\beta- \rm{and}~\gamma-$ events) measured for a 151.7 days.
The most prominent gamma lines are produced by $^{207}\rm{Bi}$ decay.} \label{Fig2}
\end{figure}

\section{RESULTS}
The detector was operated for a total live time of 151.7 days. During the measurement, various calibration runs were
performed in order to monitor the stability of the detector. By means of the different LY of interacting particles we
were able to discriminate $\alpha-$events from $\beta/\gamma$ one. This allowed us to strongly increase our
sensitivity, because we were able to reject all $\alpha-$events in the region of interest.  The total acquired
background statistics in the range of (0.3-4) MeV for $\beta/\gamma$ events are in shown in Fig.\ref{Fig2}.

In the energy spectrum it is possible to identify the most intense internal $\beta/\gamma$ background sources, which
are mainly ascribed to $^{207}\rm{Bi}$, produced by proton-induced reactions on $^{206}\rm{Pb}$ \cite{Lew88} and
$^{208}\rm{Bi}$. In first approximation, the energy resolution of large mass bolometric detector is independent of the
energy. Our FWHM energy resolution is $33.7\pm0.6$ keV at 2614 keV ($^{208}\rm{Tl}$) and $33.2\pm0.1$ keV at 570 keV
($^{207}\rm{Bi}$).

The limits on the 5.5 MeV axion flux and cross-section are based on the experimental fact that no events above 4 MeV
were observed. The upper limit on the number of axioelectric effects is $S_{lim}$ = 2.44 with 90\% c.l. in accordance
with the Feldman-Cousins procedure \cite{Fel98}.

The expected number of axioelectric absorption events are:
\begin{equation}
S_{abs} = \varepsilon N_{Bi}T\Phi_A\sigma_{Ae}
\end{equation}
where $\sigma_{Ae}$ is the axioelectric effect cross section, given by expression (\ref{sigmaAE}); $\Phi_A$ is the
axion flux (\ref{FluxA}); $N_{Bi} = 6.87 \times 10^{24}$ is the number of Bi atoms; $T = 1.31\times10^7$ s is the
measurement time; and $\varepsilon = 0.59$ is the detection efficiency for 5.5 MeV electrons. The axion flux $\Phi_A$
is proportional to the constant $(g^3_{AN})^2$, and the cross section $\sigma_{Ae}$ is proportional to the constant
$g^2_{Ae}$ , according to expressions (\ref{FluxA}) and (\ref{sigmaAE}). As a result, the $S_{abs}$ value depends on
the product of the axion-electron and axion-nucleon coupling constants: $(g_{Ae})^2\times (g^3_{AN})^2$.
\begin{figure}
\includegraphics[width=9cm,height=10.5cm]{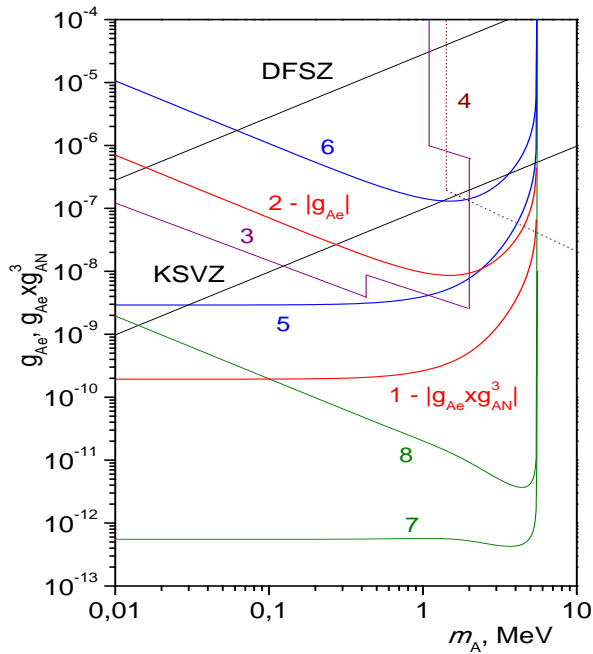}
\caption{The limits on the $g_{Ae}$ coupling constant obtained by 1,2 - present work for $|g_{Ae}|$ and $|g_{Ae}\times
g^3_{AN}|$,  correspondingly; 3- solar \cite{Bel08} and reactor experiments \cite{Alt95,Cha07}, 4- beam dump
experiments \cite{Kon86,Bjo88}; 5,6 - limits on $|g_{Ae}|$ and $|g_{Ae}\times g^3_{AN}|$ \cite{Der13}; 7,8 - Borexino
results for $|g_{Ae}|$ and $|g_{Ae}\times g^3_{AN}|$ \cite{Bel12}.  The relations between $g_{Ae}$ and $m_A$ for DFSZ-
and KSVZ-models are also shown .} \label{Fig3}
\end{figure}
The experimentally found condition $S_{abs} \leq S_{lim}$ imposes some constraints on the range  of  possible
$|g_{Ae}\times g^3_{AN}|$  and $m_A$ values. The range of excluded $|g_{Ae}\times g^3_{AN}|$ values is shown in
Fig.~\ref{Fig3}, at $m_A \rightarrow 0$ the limit is
\begin{equation}
|g_{Ae}\times g^3_{AN}| \leq 1.9\times10^{-10}~~\rm{at~90\%~c.l.}. \label{limgaegan}
\end{equation}

The dependence of $|g_{Ae}\times g^3_{AN}|$ on $m_A$ is related only to the kinematic factor in formulae (\ref{ratio})
and (\ref{sigmaAE}). These constraints are completely model-independent and valid for any pseudoscalar particle with
coupling $|g_{Ae}|$ less than $10^{-6(4)}$.

Within the hadronic axion model, $g_{AN}^3$  and $m_A$ quantities are related by expression (\ref{gan3}), which can be
used to obtain a constraint on the $g_{Ae}$ constant, depending on the axion mass (Fig.~\ref{Fig3}). For $m_A$ = 1 MeV,
this constraint corresponds to $|g_{Ae}|\leq 9.6 \times 10^{-9}$ at $\rm{90\% c.l.}$. (see Fig.~\ref{Fig3}).

These limits are more than one order of magnitude stronger than ones obtained with the 2.5 kg BGO scintillation
detector \cite{Der13}. Figure \ref{Fig3} also shows the constraints on the constant $|g_{Ae}|$ that were obtained in
the Borexino experiment for 478 keV ${^7\rm{Li}}$ solar axions \cite{Bel08} and in the Texono reactor experiment  for
2.2 MeV axions produced in the $n + p \rightarrow d + A$ reaction \cite{Cha07}. Recently, Borexino coll. reported new
more stringent limits on $g_{Ae}$ coupling for 5.5 MeV solar axions \cite{Bel12}. Unlike our work, these limits on
$g_{Ae}$ were obtained in assumption that the axion interacts with electron through the Compton conversion process. The
cross section of axioelectric effect has a $Z^5$ dependence (\ref{sigmaAE}) and for carbon atoms (the main component of
liquid scintillator) the cross section is $5\times10^5$ times lower than for bismuth ones.

In the model of the mirror axion \cite{Ber01} an allowed parameter window is found within the P-Q scale
$f_A\sim10^4-10^5$ GeV and the axion mass $m_A\sim$ 1 MeV. The limit (\ref{limgaegan}) on the product $|g_{AN}^3\times
g_{Ae}| \leq 1.9\times 10^{-10}$ may be represented as a limit on the value $f_A$ by taking the following relations
into account: $g_{AN}^3=0.5(g_{Ap}-g_{An})= 1.1/f_A$ and $g_{Ae}=5\times10^{-4}/f_A$. For axion masses about 1 MeV
discussed in \cite{Ber01}, the limit is $f_A > 1.7\times10^{3} \rm{~GeV}$, which is close to the lower bound of  mirror
axion window.


Our results set constraints on the parameter space of the CP-odd Higgs $(A^0)$, which arise in the next-to-minimal
supersymmetric Standard Model due to the spontaneous breaking of approximate symmetries such as PQ-symmetry, and is
motivated by the string theory \cite{Hoo09,And10}. The corresponding exclusion region can be obtained from
Fig.\ref{Fig3} using the conversion $C_{Aff} = g_{Ae}2m_W/g_2 m_e$ where $C_{Aff}$ is the coupling of the CP-odd Higgs
to fermions and $g_2=0.62$ is the gauge coupling. Taking the relation (\ref{gan3}) into account, the limit
((\ref{limgaegan})) translates into $C_{Aff}\times m_{A^0} \leq 3\times 10^{-3}$ MeV for $m_{A^0} < 1$ MeV, which is
compatible with the limits obtained in reactor experiments exploring Compton conversion.

\section{CONCLUSIONS}
A search for the axioelectric absorption of 5.5 MeV axions  produced in the $p + d \rightarrow {^3{\rm{He}}} + \gamma$
reaction was performed using four BGO bolometric detectors with a total mass of 3.56 kg, located in a low-background
setup equipped with passive and active shielding. As a result, a model-independent limit on axion-nucleon and
axion-electron coupling constant has  been obtained: $| g_{Ae}\times g_{AN}^3|< 1.9\times 10^{-10}$ (90\% c.l.). Within
the hadronic axion model the constraints on the axion-electron coupling constant $|g_{Ae}|\leq (9.6 - 82) \times
10^{-9}$ for axions with masses $0.1 < m_A < 1$ MeV were obtained for 90\% c.l.. The obtained constraints are compared
with the parameter space of the mirror axions and light CP--odd Higgs models.

\section{ACKNOWLEDGMENTS}
This work was supported by RFBR grants 13-02-01199 and 13-02-12140-ofi-m.

\end{document}